\documentclass[aps,prd,10pt,nofootinbib,twocolumn]{revtex4-1}
\usepackage{amsmath,amssymb,amsfonts,dsfont,mathrsfs,amsthm,mathtools}
\usepackage{hyperref}
\usepackage[english]{babel}
\hypersetup{linktocpage,colorlinks=true,urlcolor=blue,linkcolor=blue,citecolor=blue}
\usepackage[pdftex]{graphicx}
\usepackage{pgf,tikz,wasysym}
\usepackage{tensor}

\begin{document}

\title{Unimodular gravity as an initial value problem}

\author{J.E. Herrera}
\author{Y. Bonder}
\email{bonder@nucleares.unam.mx}
\affiliation{Instituto de Ciencias Nucleares, Universidad Nacional Aut\'onoma 
de M\'exico\\ Apartado Postal 70-543, Cd.~Mx., 04510, M\'exico}

\begin{abstract}
Unimodular gravity is a compelling modified theory of gravity that offers a natural solution to the cosmological constant problem. However, for unimodular gravity to be considered a viable theory of gravity, one has to show that it has a well-posed initial value formulation. Working in vacuum, we apply Dirac's algorithm to find all the constraints of the theory. Then, we prove that, for initial data compatible with these constraints, the evolution is well-posed. Finally, we find sufficient conditions for a matter action to preserve the well-posedness of the initial value problem of unimodular gravity. As a corollary, we argue that the ``unimodular'' restriction on the spacetime volume element can be satisfied by a suitable choice of the lapse function.
\end{abstract}

\maketitle

\section{Introduction}

Nowadays, general relativity (GR) is accepted as the theory of gravity. However, GR is not problem-free. In particular, it requires a cosmological constant to describe the universe at cosmological scales and its measured value departs, by many orders of magnitude, from the value estimated by considering vacuum state contributions \cite{Weinberg}. Remarkably, there is a modified theory of gravity, called unimodular gravity (UG), where the cosmological constant arises as an integration constant, and it is thus independent of vacuum state contributions \cite{Percacci}.

One important property of any fundamental physical theory is its ability to predict a system's evolution from initial data. Given some initial data, perhaps subject to some constraints, this evolution ought to be unique. However, for some theories, these properties are not enough; the evolution must also be continuous and causal in the following sense: we expect that small perturbations on initial data should produce small changes in the solutions, where the notion of ``smallness'' is given by certain Sobolev norms \cite{Wald}. In other words, we require the solutions to depend continuously on the initial data to avoid losing predictability, since the initial data can only be measured with finite precision. Also, changes in initial data supported in a given spacetime region should only affect the region's causal future (and past). A relativistic theory in which the evolution is unique, continuous, and causal, in the above-described sense, is said to have a well-posed initial value formulation.

A very important feature of GR is that it has a well-posed initial value formulation \cite{Choquet-Bruhat1969}. This is not a trivial result since the metric, which is the dynamical field, contains all the causal information. Nevertheless, the well-posedness of the initial problem of GR can be shown by writing the equation of motion, via a judicious choice of coordinates, in a form where one can prove that the above-mentioned properties hold.

It is worth mentioning that, besides GR, there is a relatively small set of modified gravity theories for which proofs of well-posed initial value formulation are known. Examples of modified gravity theories where such results have been obtained include scalar-tensor theories \cite{Salgado}, the $k$-essence theory \cite{Rendall}, the Einstein-{\ae}ther theory \cite{Sarbachetal}, Horndeski theories \cite{Kovacs, KYR}, and a $4$-derivative scalar-tensor theory \cite{Salo}. Still, to the best of our knowledge, there are no previous proofs of this kind for theories with nondynamical tensors.

The goal of this work is to show that UG has a well-posed initial value formulation. We must stress that the proof we present is not a simple application of the GR techniques. This is because the UG constraint structure is different from that of GR. In this sense, this work introduces methods that could be used when investigating the initial value formulation of other modified gravity theories.

We structured the paper as follows: in Sec.~\ref{preliminaries} we introduce UG. Sec.~\ref{sec. 4.2} is the core of this paper and it begins with a classification of the equations of motion of vacuum UG into evolution equations and (primary) constraint. Then, we find all the constraints necessary for a consistent evolution (Subsec.~\ref{sec constraints}), we study the evolution equations in the well-known BSSN formulation, named after Baumgarte, Shapiro, Shibata, and Nakamura (Subsec.~\ref{evolutioneqs}), and we propose sufficient conditions for UG with matter to have a well-posed initial value formulation (Subsec.~\ref{matter}). Finally, we offer our conclusions in Sec.~\ref{conclusions}. In the Appendix, we present some mathematical tools that are used throughout the paper and we clarify our notation.

\section{Basic aspects of Unimodular Gravity}\label{preliminaries}

Historically, UG dates back to Einstein \cite{Einstein} and Pauli \cite{PauliGR} who were interested in possible interplays between gravity and elementary particles (for historical remarks see Ref.~\onlinecite{EAEV}). Yet, the framework that is closer to what is presented here emerged some fifty years later \cite{AndersonF} when the theory was studied in the context of field theory \cite{Bij, Dragon1, Dragon2}. In 2011, attention was drawn to UG with the observation that the energy associated with the vacuum state does not gravitate (in a semiclassical framework), bypassing the cosmological constant problem \cite{Ellis}; a claim that is not free of criticism \cite{Earman}.

Recent works rekindled the interest in UG. For example, it has been shown that energy nonconservation avoids some incompatible features with Quantum Mechanics and could give rise to an effective cosmological constant that has an adequate sign and size \cite{Sudarsky1, Sudarsky3}. Also, cosmological diffusion models in the UG framework affect the value of the Hubble constant \cite{Nucamendi1}, among other interesting features \cite{Nucamendi2}.

Here, we work on a $4$-dimensional spacetime $M$ equipped with the pseudo-Riemannian metric $g_{ab}$ (we follow the notation and conventions of Ref.~\onlinecite{Wald} where pairs of indexes between parentheses/brackets stand for its symmetric/antisymmetric part, with a $1/2$ factor). Moreover, spacetime is assumed to be globally hyperbolic, which allows us to foliate $M$ by constant time (Cauchy) hypersurfaces $\Sigma_t$.

There are several ways to introduce UG \cite{review}; the UG action we consider here is
\begin{eqnarray}
\mathrm{S}\left[g^{ab}, \lambda, \Phi\right]&=&\frac{1}{2\kappa}\int\mathrm{d}^4 x \left\lbrace \sqrt{-g} R+\lambda\left(\sqrt{-g}-f\right) \right\rbrace\nonumber\\
&&+\mathrm{S_M}\left [g^{ab},\Phi\right],\label{action}
\end{eqnarray}
where $g^{ab}$ is the inverse of $g_{ab}$, $\lambda$ is a scalar field that acts as a Lagrangian multiplier, $\kappa$ is the gravitational coupling constant, and $R$ is the curvature scalar associated with the metric-compatible and torsion-free derivative $\nabla_a$. Moreover, $g$ is the determinant of $g_{ab}$, and $f$ is a nondynamical positive scalar density, i.e., a real function that transforms under coordinate transformations mimicking $\sqrt{-g}$. The matter action is
\begin{equation}\label{2.18}
\mathrm{S_M}\left[g^{ab},\Phi\right]= \int \mathrm{d}^4x\sqrt{-g} \mathcal{L}_\mathrm{M} \left(g^{ab},\Phi\right) ,
\end{equation}
where $\mathcal{L}_\mathrm{M}\left(g^{ab},\Phi\right)$ is the matter Lagrangian and $\Phi$ collectively describes all matter fields.

We want to emphasize that the only difference in the UG action when compared with that of GR is the presence of the term with the Lagrange multiplier. This term fixes the differential spacetime volume element, $\sqrt{-g}\mathrm{d}^4 x$, to coincide with $f\mathrm{d}^4 x$, which requires $f>0$.

An arbitrary variation of the action \eqref{action} has the form
\begin{eqnarray}
\delta\mathrm{S}&=&\int\mathrm{d}^4 x \left\lbrace \sqrt{-g} \left[ \frac{1}{2\kappa}(G_{ab}-\frac{1}{2}g_{ab}\lambda)-\frac{1}{2}T_{ab}\right]\delta g^{ab}\right.\nonumber\\
&&+\left.\frac{1}{2\kappa}\left(\sqrt{-g}-f\right)\delta\lambda+\frac{\delta \mathcal{L}_\mathrm{M} }{\delta \Phi}\delta \Phi\right\rbrace,\label{actionvar}
\end{eqnarray}
where we omit the boundary terms, as we do throughout the paper, and we define the energy-momentum tensor
\begin{equation}\label{2.20}
T_{ab}\coloneqq -\frac{2}{\sqrt{-g}}\frac{\delta\left(\mathcal{L}_\mathrm{M}\sqrt{-g}\right)}{\delta g^{ ab}}.
\end{equation}
Hence, the metric equation of motion is
\begin{equation}\label{2.21}
R_{ab}-\frac{1}{2}R g_{ab}-\frac{1}{2}\lambda g_{ab}=\kappa T_{ab}.
\end{equation}
Furthermore, the equation of motion associated with $\lambda$ yields the ``unimodular constraint''
\begin{equation}\label{unimodularconstraint}
\sqrt{-g}=f.
\end{equation}
Of course, there are also matter field equations that can be written generically as
\begin{equation}\label{MatterEOM}
\frac{\delta \mathrm{S_M}}{\delta \Phi} =0.
\end{equation}

The divergence of Eq.~\eqref{2.21} produces
\begin{equation}\label{2.26}
\kappa\nabla^a T_{ab}=-\frac{1}{2}\nabla_b \lambda,
\end{equation}
where we use the Bianchi identity. Interestingly, if $\nabla^a T_{ab}=0$, $\lambda$ is constant, and Eq.~\eqref{2.21} becomes the conventional Einstein equation with $\lambda$ playing the role of a cosmological constant. Thus, vacuum UG, in particular, is equivalent, at the level of the equations of motion, to GR with a cosmological constant. Still, UG is compatible with some matter solutions where $\nabla^a T_{ab}\neq 0$, opening the door to new phenomenology \cite{BonderHerreraRubiol}.

On the other hand, the trace of Eq.~\eqref{2.21} produces
\begin{equation}\label{2.22}
\lambda =-\frac{1}{2}(R+\kappa T),
\end{equation}
where $T\coloneqq g^{ab} T_{ab}$. Inserting Eq.~\eqref{2.22} in Eq.~\eqref{2.21} yields
\begin{equation}\label{2.23}
E_{ab}\coloneqq  R_{ab}-\frac{1}{4}Rg_{ab}-\kappa \left(T_{ab}-\frac{1}{4} T g_{ab}\right)=0,
\end{equation}
which satisfies
\begin{equation}\label{traceless}
E_{ab}g^{ab}=0.
\end{equation}

Another difference between GR and UG is the theories' symmetries and the corresponding conservation laws. As it is well known, GR is invariant under all diffeomorphisms, which implies that $\nabla_a T^{ab}=0$. This is not the case in UG where the nondynamical function $f$, which does not transform under (active) diffeomorphisms, partially breaks invariance under diffeomorphisms \cite{Cristobal1}. To show this, we first consider theory in vacuum, namely, $\mathrm{S_M}=0$. The variation of the vacuum UG action with respect to a diffeomorphism associated with the vector field $\xi^a$ is given by Eq.~\eqref{actionvar} with
\begin{eqnarray}
 \delta g^{ab}&=&\pounds_\xi g^{ab}=-2 \nabla^{(a} \xi^{b)},\\ 
 \delta \lambda &=&\pounds_\xi \lambda=\xi^a \nabla_a \lambda,
\end{eqnarray}
where $\pounds_\xi$ is the Lie derivative along $\xi^a$. Then, the UG action variation with respect to a diffeomorphism takes the form
\begin{eqnarray}
\delta\mathrm{S}&=&\int\mathrm{d}^4 x \left\lbrace \left[ \frac{1}{\kappa}(-G_{ab}+\frac{1}{2}g_{ab}\lambda)\right]\nabla^{a} \xi^{b}\right.\nonumber\\
&&+\left.\frac{1}{2\kappa}\left(\sqrt{-g}-f\right)\xi^a \nabla_a \lambda\right\rbrace.\label{actionvardiff}
\end{eqnarray}
After we integrate by parts (with the appropriate volume form), this variation can be written as
\begin{equation}
\delta\mathrm{S}=\frac{1}{2\kappa}\int\mathrm{d}^4 x \sqrt{-g}\lambda \nabla_a\left( \xi^{a} \frac{f}{\sqrt{-g}}\right),\label{actionvardiff2}
\end{equation}
where we use the Bianchi identity. On shell, Eq.~\eqref{unimodularconstraint} is valid, and thus,
\begin{equation}
\delta\mathrm{S}=\frac{1}{2\kappa}\int\mathrm{d}^4 x \sqrt{-g}\lambda \nabla_a \xi^{a}.\label{actionvardiff3}
\end{equation}
Hence, the vacuum UG action is only invariant under diffeomorphisms associated with divergence-free vector fields. We refer to this restricted set of diffeomorphisms as volume-preserving diffeomorphisms.

To obtain the matter conservation law associated with volume-preserving diffeomorphisms, we notice that a divergence-free vector field $\xi^a$ can be written in terms of a \emph{generic} antisymmetric tensor $\alpha_{ab}$ as
$\xi^a = \epsilon^{abcd}\nabla_b\alpha_{cd}$, where $\epsilon_{abcd}$ is the volume form associated with $g_{ab}$. Thus, the on-shell variation of $\mathrm{S_M}$ with respect to a volume-preserving diffeomorphism can be written as
\begin{eqnarray}
\delta\mathrm{S_M}&=&\int\mathrm{d}^4 x \sqrt{-g} T_{ab}\epsilon^{bcde}\nabla^{a} \nabla_c\alpha_{de}\nonumber\\
&=& \int d^4 x \sqrt{-g}\alpha_{de} \epsilon^{bcde}\nabla_c \nabla^{a} T_{ab} ,
\end{eqnarray}
where, in the last step, we integrate by parts twice. Therefore, the matter action is invariant under all volume-preserving diffeomorphisms if
\begin{eqnarray}
\nabla_{[b} \nabla^{a} T_{c]a}=0.
\end{eqnarray}
This last equation is the UG matter conservation law, and it is more general than $\nabla^a T_{ab}=0$, which is the matter conservation law of GR. What is more, using the Poincar\'e Lemma \cite{Nakahara:2016} and under the hypothesis that $M$ is simply connected, this conservation law implies that there exists a scalar $Q$ such that
\begin{eqnarray} \label{nablaQ}
\nabla^{a} T_{ab}=\nabla_b Q.
\end{eqnarray}
Notice that, when $Q$ is constant, we recover the GR matter conservation law. However, in UG, $Q$ can be arbitrary (still, people have considered $T_{ab}-g_{ab} Q$ as an effective energy-momentum tensor that is conserved \cite{Percacci}). These are all the UG aspects required in this paper, and we turn now to classify the UG equations of motion as evolution or constraints and study the Cauchy problem for UG.

\section{Initial value problem}\label{sec. 4.2}

In this section, we study UG in terms of evolving $3$-dimensional geometrical objects. We recommend to a reader who is not familiar with the mathematical tools of the $3+1$ decomposition of GR, or with the corresponding notation, to review the Appendix.

The first task when studying the initial value problem of a theory is to identify the constraints. When working with a geometrical gravity theory that has second-order equations of motion, an evolution equation is, by definition, an equation of motion that contains time derivatives of the extrinsic curvature. This is because the time derivatives can be thought of as second-time derivatives of $h_{ab}$ [see Eq.~\eqref{4.19}]. Conversely, an equation with no time derivatives of $K_{ab}$ is a constraint, which must be imposed on the initial data, and, for consistency, must be kept valid under evolution.

Notice that, to perform an initial value study, we need to specify the matter action. Thus, before we incorporate matter fields, we consider vacuum UG and we later use the lessons from vacuum UG to analyze UG with matter. We define $E^{\rm (v)}_{ab}\coloneqq  R_{ab}-g_{ab}R/4$, which coincides with $E_{ab}$ when $T_{ab}=0$. Importantly, in this part of our study, $E^{\rm (v)}_{ab}$ is \emph{not} assumed to be zero throughout $M$. Instead, a weaker assumption is considered: the constraints are only valid on the initial data hypersurface, while the evolution equations are satisfied all over $M$. Note that, since we do not assume that all the equations of motion are valid throughout $M$, their $4$-dimensional divergences are not well defined.

Of course, we expect vacuum UG to have a well-posed initial value formulation, since this theory is dynamically equivalent to GR with a cosmological constant. Still, this is not an obvious result since we cannot take $4$-dimensional divergences of the equations of motion, which is the key step to show the equivalence between UG and GR with a cosmological constant.

Using the Bianchi identity, we can show that
\begin{equation}
\nabla^a E^{\rm (v)}_{ab}=\frac{1}{4}\nabla_b R.\label{Bianchi E}
\end{equation}
However, under the assumptions we consider, we cannot claim that its divergence vanishes, and thus, at this stage we \emph{cannot} argue that $R$ is constant. Also, observe that, from Eq.~\eqref{traceless}, we get
\begin{equation}
h^{ab}E^{\rm (v)}_{ab}=E^{\rm (v)}_{ab} n^a n^b. \label{trace E}
\end{equation}
This last equation allows us to identify the $3$-dimensional trace of tangential-tangential projection with the normal-normal projection, rendering the latter redundant; in what follows, we omit the normal-normal projection. Thus, there are only nine independent equations of motion, which, together with the unimodular constraint, amount to ten equations, which coincides with the number of independent equations of GR.

Without loss of generality, we take the initial value hypersurface to be $\Sigma_0$. Since we assume all equations of motion ---evolution equations and constraints--- to vanish on the initial value hypersurface, we impose $E^{\rm (v)}_{ab}|_{\Sigma_0}=0$. We then separate $E^{\rm (v)}_{ab}|_{\Sigma_0}=0$ into its tangential-tangential projection and its normal-tangential projection, which are, respectively, given by
\begin{eqnarray}
0 &=& {^{\left(3\right)}R_{ab}} +KK_{ab} -2 \tensor{K}{_a^c}K_{cb}+N^{-1}\dot{ K}_{ab} -a_a a_b \nonumber \\
& &-N^{-1}N^c D_c K_{ab} -2N^{-1}K_{c( a}D_{b )}N ^c -D_{( a} a_{b)} \nonumber \\
& & -\frac{1}{4} \big[ {^{\left(3\right)}}R +K^2 -3K_{cd}K^{cd} +2N^{-1} h^ {cd}\dot{K}_{cd} \nonumber \\
&& -2N^{-1} N^c D_c K -4N^{-1} K^{cd} D_c N_d -2a_c a^c \nonumber \\
& &-2D_c a^c \big] h_{ab}, \label{4.38}\\
0 &=& D_b \tensor{K}{^b_a}-D_a K,\label{4.41}
\end{eqnarray}
where we use some results from the Appendix, concretely Eqs.~\eqref{Rhh} and \eqref{Rnh}. The tangential-tangential projection, Eq.~\eqref{4.38}, has time derivatives of $K_{ab}$. Therefore, it is an evolution equation. On the other hand, the normal-tangential projection, Eq.~\eqref{4.41}, does not contain time derivatives of $K_{ab}$ and is thus a constraint. This constraint is reminiscent of the GR momentum constraint.

Notice that the unimodular constraint, Eq.~\eqref{unimodularconstraint}, is a constraint in the sense that it does not have time derivatives of $K_{ab}$. Still, it is not a relation that can be imposed on the initial data and be automatically satisfied throughout spacetime by the evolution. This is because the values of $f$, which is nondynamical, are given \textit{a priori} all over $M$. However, we will show that the unimodular constraint can be satisfied by choosing the lapse function.

In summary, in terms of components, six of the field equations of vacuum UG, the tangent-tangent projections, are evolution equations, while the remaining three, the normal-tangential projections, are constraints. In the next subsection, we analyze the evolution of the constraints.

\subsection{Constraint equations}\label{sec constraints}

We now study whether the constraint is maintained under evolution. In other words, given initial data on $\Sigma_0$ that satisfy Eq.~\eqref{4.41}, we must verify if the fields obtained by evolving this initial data still satisfy the constraint. This can be done by proving that the constraint, on $\Sigma_0$, has zero time derivative. 

We define the tangential tensors
\begin{eqnarray}
    \mathcal{E}_{ab} &\coloneqq& h_a^c h_b^d E^{\rm (v)}_{cd},\\
C_a &\coloneqq& D_b \tensor{K}{^b_a}-D_a K .\label{4.42}
\end{eqnarray}
Let $A_1$ be the following set of assumptions: the evolution equations, $\mathcal{E}_{ab} =0$, are valid throughout $M$, and the constraint, $C_a=0$, is only valid on $\Sigma_0$. What remains to check is if, under $A_1$, $\dot{C}_a=0$ on $\Sigma_0$. From the definition of the time derivative (see the Appendix), we readily obtain
\begin{eqnarray}
\dot{C}_a &=& N h_a^d n^e \nabla_e C_d +N^e D_e C_a \nonumber\\
&&+N C_e \tensor{K}{^e_a}+ C_e D_a N^e.
 \label{4.46}
\end{eqnarray}
Given that $ C_a=0=D_a C_b$ on $\Sigma_0$, Eq.~\eqref{4.46}, when restricted to $\Sigma_0$, takes the form
\begin{equation}\label{4.50}
\left.\dot{C}_a\right|_{\Sigma_0} = N h_a^b n^c \nabla_c C_b ,
\end{equation}
which is \emph{not} automatically zero under $A_1$.

Let
\begin{equation}\label{4.51}
S_a \coloneqq h_a^b n^c\nabla_c C_b.
\end{equation}
The time derivative of $C_a$ vanishes if $S_a=0$. We can show that, under $A_1$, $S_a=-D_a R/4$ on $\Sigma_0$: 
\begin{proof}
We can verify that
\begin{equation}
h_a^d\nabla^b\left(E^{\rm (v)}_{bc} h^c_d\right) 
= D^b \mathcal{E}_{ab} + \mathcal{E}_{ab} a^b - S_a - KC_a ,\label{4.54}
\end{equation}
where we use $K = \nabla_c n^c$. Now, using Eq.~\eqref{Bianchi E}, we obtain
\begin{equation}
h_a^d\nabla^b\left(E^{\rm (v)}_{bc} h^c_d\right) 
=\frac{1}{4}D_a R +  \tensor{K}{_a^b}C_b - a_a \mathcal{E}, \label{4.59}
\end{equation}
where we define $a^a \coloneqq n^b\nabla_b n^a$ and
\begin{eqnarray}
\mathcal{E}&\coloneqq& \mathcal{E}_{ab}h^{ab}\nonumber\\
&=&\frac{1}{4} {^{\left(3\right)}R} +\frac{1}{4}K^2 +\frac{1}{4} K_{ab} K^{ab} -\frac{1}{2} N^{-1}h^{ab}\dot{ K}_{ab}  \nonumber \\
& &+ \frac{1}{2}a_a a^a+\frac{1}{2} N^{-1} N^c D_c K + N^{-1}K^{ab} D_a N_b \nonumber \\
&&+ \frac{1}{2}D_a a^a.\label{E}
\end{eqnarray}
Combining Eqs.~\eqref{4.54} and \eqref{4.59} produces
\begin{eqnarray}
 S_a &=&-\frac{1}{4}D_a R - \tensor{K}{_a^b}C_b - KC_a+ \mathcal{E}_{ab} a^b\nonumber\\
&& + a_a \mathcal{E} +D^b \mathcal{E}_{ab}.\label{S}
\end{eqnarray}
Hence, on $\Sigma_0$ and assuming $A_1$,
\begin{eqnarray}
\left.S_a \right|_{\Sigma_0}=-\frac{1}{4}\left.D_a R\right|_{\Sigma_0} .\end{eqnarray}
\end{proof}

According to Dirac's method \cite{Dirac}, it is necessary to promote $S_a=0$ as a constraint; following Dirac's terminology, $S_a=0$ is a ``secondary constraint.'' This, of course, implies that the $4$-dimensional curvature scalar, $R$, must be constant throughout $\Sigma_0$. Namely, $R=4 \Lambda$, where $D_a \Lambda=0$. Notice that we must consider $R$ as a shorthand notation for the right-hand side of Eq.~\eqref{RGC}, which only contains tangential objects. Also, as the notation suggests, $\Lambda$ will end up playing the role of the cosmological constant. However, at this stage, we can only claim that $\Lambda$ is constant along $\Sigma_0$ and there is \emph{no} reason to assume that $\dot{\Lambda}=0$. Surprisingly, only when we require dynamical consistency, do we find that $\Lambda$ is constant throughout $M$.

We now prove that, under $A_1$, $\dot{R}=0$ on $\Sigma_0$:
\begin{proof}
Using the fact that $h_a^b-n_a n^b$ is the identity tensor in spacetime, we can verify that
\begin{equation}
E^{\rm (v)}_{ab} n^b=C_a- n_a \mathcal{E}.
\end{equation}
The divergence of this last equation produces
\begin{equation}\label{4.68p}
\nabla^a (E^{\rm (v)}_{ab} n^b)= \nabla_a C^a - K \mathcal{E}- n^a \nabla_a\mathcal{E}.\end{equation}
Alternatively, we can calculate $\nabla^a (E^{\rm (v)}_{ab} n^b)$ using the Leibniz rule and Eq.~\eqref{Bianchi E}, which yields
\begin{equation}
\nabla^a\left( E^{\rm (v)}_{ab} n^b\right) =\frac{1}{4}n^a \nabla_a R + E^{\rm (v)}_{ab} K^{ab} - C_a a^a.\label{4.68}
\end{equation}
When we compare Eqs.~\eqref{4.68p} and \eqref{4.68}, we obtain
\begin{equation}\label{39}
D_a {C}^a + 2C_a a^a - K \mathcal{E}- n^a \nabla_a\mathcal{E} =\frac{1}{4}n^a \nabla_a R+ \mathcal{E}_{ab} K^{ab},
 \end{equation}
where we use $\nabla_a {C}^a = D_a {C}^a + C_a a^a$. Hence, on $\Sigma_0$ and assuming $A_1$,
\begin{equation}
\left.n^a \nabla_a R\right|_{\Sigma_0}=0.\label{4.68b}
\end{equation}
This result, together with the fact that $D_a R=0$ on $\Sigma_0$, implies that $\dot{R}=0$ on $\Sigma_0$.
\end{proof}

The lesson from this last proof is that, under $A_1$, $\Lambda$, which coincides with $R/4$ on $\Sigma_0$, is constant throughout $M$. Hence, $\Lambda$ can play the role of the cosmological constant, as we anticipated. Recall that $A_1$ is a significantly weaker set of assumptions than assuming that $E^{\rm (v)}_{ab}=0$ throughout $M$.

We now need to check if $\dot{S}_a=0$ on $\Sigma_0$, assuming $A_1$ and $\left.S_a \right|_{\Sigma_0}=0$; we refer to this new set of assumptions by $A_2$. Clearly, under $A_2$, the vanishing of $\dot{S}_a$ is equivalent to
\begin{equation}
\left.h_a^b n^c\nabla_c S_b\right|_{\Sigma_0}=0,\label{Sdot}
\end{equation}
which we show to hold:
\begin{proof}
Using Eq.~\eqref{S}, we get
\begin{eqnarray}
h_a^b n^c\nabla_c S_b
&=&
 -\frac{1}{4}h_a^b n^c\nabla_c D_b R -C_d h_a^b n^c\nabla_c \tensor{K}{_b^d} \nonumber\\
 &&- \tensor{K}{_a^b} S_b -C_a n^b\nabla_b K -K S_a \nonumber\\
 && +a^d h_a^b n^c\nabla_c \mathcal{E}_{bd}+ \mathcal{E}_{ab} n^c\nabla_c a^b+ a_a n^b\nabla_b \mathcal{E} \nonumber\\
 && + \mathcal{E}h_a^b n^c\nabla_c a_b +h_a^b n^c\nabla_c D^d \mathcal{E}_{bd}  ,\label{eqSdot}
 \end{eqnarray}
where we write $n^c\nabla_c \tensor{K}{_a_b} $ and $n^a\nabla_a K$ in terms of $R=4 \Lambda$ and other tangential objects using Eqs.~\eqref{GC R} and \eqref{K dot}. Notice that, except for the first term, all the terms in Eq.~\eqref{eqSdot} are proportional to $C_a$, $S_a$, $\mathcal{E}$, $\mathcal{E}_{ab}$ or derivatives of $\mathcal{E}$ and $\mathcal{E}_{ab}$, all of which vanish under $A_2$. Thus, we need to focus on
\begin{eqnarray}
h_a^b n^c\nabla_c D_b R&=&h_a^b n^c\nabla_c (h_b^d \nabla_d R)\nonumber\\
&=&h_a^b (n^c\nabla_c h_b^d) \nabla_d R+h_a^b n^c\nabla_c \nabla_b R \nonumber\\
&=&a_a n^b \nabla_b R+D_a(n^b \nabla_b R) - \tensor{K}{_a^b} D_b R,\nonumber\\
&&\label{42}
\end{eqnarray}
where we repeatedly use the definition of the tangential derivative, the Leibniz rule, and the fact that covariant derivatives acting on scalars commute; this is a consequence of the torsion-free hypothesis (a study of a unimodular theory with torsion is presented in Ref.~\onlinecite{PhysRevD.97.084001}). When we evaluate on $\Sigma_0$, we see that the first and last terms in Eq.~\eqref{42} vanish using $R=4\Lambda$. Moreover, if we take the tangential derivative of Eq.~\eqref{39}, we can write the second term in terms of objects that also vanish under $A_2$. With all this, we can show that, under $A_2$, $h_a^b n^c\nabla_c S_b=0$ on $\Sigma_0$.
\end{proof}

Relevantly, the secondary constraint $S_a=0$ can be written more familiarly. Starting from $R= 4\Lambda$, which, under $A_2$, is equivalent to $S_a=0$, and using $\mathcal{E}=0$ and Eqs.~\eqref{E} and \eqref{RGC}, it is possible to obtain
\begin{equation}\label{constricHam}
4\Lambda = 2 {^{\left(3\right)}}R +2K^2-2K_{cd}K^{cd},
\end{equation}
which has the form of the Hamiltonian constraint of GR with a cosmological constant. Remarkably, this shows that vacuum UG and GR with a cosmological constant are equivalent, as expected. However, in the light of the $3+1$ formalism, this equivalence is not due to the Bianchi identity and a spacetime divergence; it arises by requiring the constraints to be maintained under evolution. Moreover, notice that, in this approach, the value of $\Lambda$ is determined by the initial data through Eq.~\eqref{constricHam}.

The main conclusion of this subsection is that, for consistency, initial data in vacuum UG must be subject to Eqs.~\eqref{4.41} and \eqref{constricHam}. In the next subsection, we verify that the evolution predicted in vacuum UG is unique, continuous, and causal in the above-described sense.

\subsection{Evolution equations}\label{evolutioneqs}

In this subsection, we study the evolution equations, namely, the tangential-tangential projection of the vacuum field equations, and the unimodular constraint. Up to this point, we have taken $h_{ab}$ and $K_{ab}$ as the dynamic variables for vacuum UG. However, we use the BSSN formulation, developed for GR initially by Shibata and Nakamura \cite{SN} and later by Baumgarte and Shapiro \cite{BS}, which is better suited to this part of our study. The BSSN formulation consists of separating the conformal factor for the spatial metric and the trace of the extrinsic curvature, and studying their evolution separately. Also, we decompose the tensors into their components on the foliation hypersurfaces, which are denoted with Latin indexes $i,j,k$. Moreover, for simplicity, we take $N^a=0$, a condition that can be trivially relaxed.

Let $ \tilde{h}_{ij}$ be such that
\begin{equation}\label{4.85}
h_{ij}= \psi^4 \tilde{h}_{ij},
\end{equation}
where $\psi= h^{1/12}$, so that the determinant of $\tilde{h}_{ij}$ is $\tilde{h}=1$ ($h$ is the determinant associated with $h_{ij}$). Notice that
\begin{equation}\label{4.88}
\tilde{h}^{ij}= \psi^{4} h^{ij}= h^{1/3} h^{ij}
\end{equation}
is the inverse of $\tilde{h}_{ij}$. Let $A_{ij}$ be the symmetric tensor field defined by
\begin{equation}\label{4.89}
A_{ij} \coloneqq K_{ij} -\frac{1}{3}K h_{ij}.
\end{equation}
Clearly $A_{ij}\tilde{h}^{ij}=0$, so that $A_{ij}$ is the traceless part of the extrinsic curvature.

In the BSSN formulation, the dynamical variables are $\phi$, $K$, $\tilde {h}_{ab}$, $\tilde{A}_{ab}$, and $\tilde{\Gamma}^a$, which are given by
\begin{subequations}\label{4.90}
\begin{eqnarray}\label{4.90a}
\phi &\coloneqq &\ln \psi = \frac{1}{12}\ln h,\\
\label{4.90b}
K&=& h^{ij} K_{ij},\\
\label{4.90c}
\tilde{h}_{ij}& =& \mathrm{e}^{-4\phi} h_{ij},\\
\label{4.90d}
\tilde{A}_{ij} &=&\mathrm{e}^{-4\phi}A_{ij},\\
\label{4.90e}
\tilde{\Gamma}^i &\coloneqq& \tilde{h}^{jk} \ \tensor{\tilde{\Gamma}}{^i_j_k}.
\end{eqnarray}
\end{subequations}
The first expression is a redefinition of the conformal factor, the second is the trace of the extrinsic curvature, the third is the conformal transformation given in Eq.~\eqref{4.85}, the fourth is a rescaling of the traceless part of the extrinsic curvature, and finally, $\tilde {\Gamma}^i$ are the conformal connection functions, where
\begin{equation}\label{4.91}
\tensor{\tilde{\Gamma}}{^i_j_k} =\frac{1}{2}\tilde{h}^{il}\left(\partial_j \tilde{h}_{kl}+\partial_k \tilde{h}_{lj}-\partial_l \tilde{h}_{jk}\right)
\end{equation}
are the Christoffel symbols associated with $\tilde{h}_{ij}$. Furthermore, we can show that Eq.~\eqref{4.90e} is equivalent to $ \tilde{\Gamma}^i= -\partial_j \tilde{h}^{ji}$. What remains to be done is to rewrite the constraint and evolution equations in terms of the BSSN variables; this will allow us to argue when comparing with the GR evolution equations, that vacuum UG has a well-posed initial value formulation.

Using Eq.~\eqref{constricHam} we can cast the evolution equation, Eq.~\eqref{4.38}, as
\begin{eqnarray}
0&=&{^{\left(3\right)}}R_{ab} +KK_{ab}- 2\tensor{K}{_a^c}K_{cb}+N^{-1}\dot{ K}_{ab}\nonumber \\
& &-N^{-1}N^c D_c K_{ab} -2N^{-1}K_{c\left(a\right.}D_{\left. b\right)}N^c\nonumber\\
&&- N^{-1}D_a D_b N -\Lambda h_{ab} ,\label{4.134}
\end{eqnarray}
which coincides with the evolution equation of GR with a cosmological constant. Thus, the system of evolution equations of vacuum UG, in BSSN variables, takes the same form as the corresponding GR equations, namely,
\begin{subequations}
\begin{eqnarray}\label{4.135a}
\partial_t \tilde{h}_{ij}&=&2N\tilde{A}_{ij},\\
\label{4.135b}
\partial_t\phi &=&\frac{1}{6}NK,\\
\label{4.135c}
\partial_t\tilde{A}_{ij}&=&\mathrm{e}^{-4\phi}\left(D_i D_j N- N {^{\left(3\right)}}R_{ij}\right )^\mathrm{TL} \nonumber\\
&&-N\left(K\tilde{A}_{ij}-2\tensor{\tilde{A}}{_i^k}\tilde{A}_{kj}\right) ,\\
\label{4.135d}
\partial_t K&=& D_k D^k N-N\left(\tilde{A}_{ij}\tilde{A}^{ij}+\frac{1}{3}K^2-\Lambda\right),\\
\label{4.135e}
\partial_t \tilde{\Gamma}^i&=& -2N\left(\tensor{\tilde{\Gamma}}{^i_j_k}\tilde{A}^{ jk} +6 \tilde{A}^{ij}\partial_j\phi -\frac{2}{3}\tilde{h}^{ij}\partial_j K \right)\nonumber\\
&&+2\tilde{A}^{ij}\partial_j N.
\end{eqnarray}
\end{subequations}
Given that GR with a cosmological constant is strongly hyperbolic \cite{Sarbach}, then, it follows from our analysis that the evolution equations of vacuum UG are also strongly hyperbolic, and thus, this theory has a well-posed evolution in the sense of being unique, continuous, and causal.

We turn now to discuss the unimodular constraint. Using the well-known expression $
\sqrt{-g}=N\sqrt{h}$ and Eq.~\eqref{4.90a}, we get $\sqrt{-g}=N\mathrm{e}^{6\phi}$. On the other hand, the unimodular constraint is $\sqrt{-g}=f$. By direct comparison, we can conclude that the unimodular constraint is satisfied as long as we take
\begin{equation}\label{4.138}
N=f\mathrm{e}^{-6\phi}.
\end{equation}
Notably, both sides of Eq.~\eqref{4.138} must be positive, avoiding a possible inconsistency. Moreover, the fact that the unimodular constraint can be solved by simply choosing the lapse function is compatible with the result of Ref.~\cite{Falta0}, where it is noted that one component of the metric is sufficient to solve the unimodular constraint.

There are cases where the initial data must be given on different charts covering $\Sigma_0$. In this case, a ``gluing'' of the different ``evolutions'' must be made. Fortunately, this procedure can be carried out in the same manner as in GR \cite{Wald}, since this procedure relies on making coordinate transformations, which are available in UG. Likewise, there is no obstruction to applying the GR proof \cite{Choquet-Bruhat1969} that shows that there is a maximal evolution of the initial data. Therefore, we can conclude that vacuum UG has a well-posed initial value formulation that gives rise to a maximal evolution of the initial data. In the next subsection, we discuss UG with matter.

\subsection{Initial value formulation with matter}\label{matter}

The initial value formulation of a UG theory with matter can only be studied once the matter action is given. Still, we can provide a set of sufficient conditions to have a well-posed initial value formulation for a UG theory with a generic matter action. Of course, the methods we present can be applied once a particular matter action is given.

The metric field equation is equivalent to $E_{ab}=0$. Again, we do \emph{not} assume that $E_{ab}$ vanishes in $M$, but only a weaker assumption: the constraints vanish on $\Sigma_0$ and the evolution equations vanish in $M$. Importantly, we can check that
\begin{equation}
\nabla^a E_{ab}=\frac{1}{4}\nabla_b \left(R -4\kappa Q + \kappa T\right) .\label{Bianchi E matter}
\end{equation}
where we use Eq.~\eqref{nablaQ}.

The tangential-tangential projection and normal-tangential projection of $E_{ab}=0$ are, respectively, given by
\begin{eqnarray}
\label{4.30 matter}
0&=&\mathcal{E}_{ab}-\kappa \left(T_{cd}h^c_a h^d_b-\frac{1}{4} T h_{ab}\right),\\
\label{4.32 matter}
0&=&C_a-\kappa T_{bc}n^b h^c_a,
\end{eqnarray}
which are tangential tensors. Recall that, since $E_{ab}g^{ab}=0$, the normal-normal projection of $E_{ab}=0$ can be obtained from Eq.~\eqref{4.30 matter}.

We know, from the vacuum UG analysis, that Eq.~\eqref{4.30 matter} is an evolution equation. On the other hand, the normal-tangential projection, Eq.~\eqref{4.32 matter}, is a constraint provided that $T_{bc}n^b h^c_a$ can be written in such a way that it does not contain second-time derivatives of $h_{ab}$ or the matter fields. Assuming this is the case, we find the conditions necessary for the constraint to remain valid under evolution. This analysis can be done as in Subsec.~\ref{sec constraints}, the main difference is that now $4\Lambda = R- 4\kappa Q+ \kappa T $ is constant, as expected from Eq.~\eqref{Bianchi E matter}. The conclusion is that to have a consistent evolution, the initial data must be subject to 
\begin{eqnarray}
D_b\tensor{K}{^b_a}-D_a K &=&\kappa T_{bc}n^b h^c_a,\ \\
{^{\left(3\right)}}R +K^2 -K_{ab}K^{ab}&=&2(\kappa T_{ab}n^a n^b+\Lambda + \kappa Q),\  \label{HamiltonConstMatter}
\end{eqnarray}
where, to write Eq.~\eqref{HamiltonConstMatter} in this form, we take the trace of Eq.~\eqref{4.30 matter} and we use $T=T_{ab}h^{ab}-T_{ab}n^an^b$. Equation \eqref{HamiltonConstMatter} looks like the Hamiltonian constraint with matter and a cosmological constant but it depends on $Q$, which, recall, is closely related to $\nabla^a T_{ab}$; this is an important difference when comparing with GR.

We can study the well-posedness of the evolution equations with matter in a simple way. Mimicking the analysis of vacuum UG in terms of BSSN variables, we can show that the UG evolution equations have the same form as those of GR with matter and a cosmological constant, but in this case $\Lambda  + \kappa T_{ab}n^a n^b+ \kappa Q$ plays the role of the cosmological constant. Now, it is known that GR with matter and a cosmological constant has a well-posed initial value formulation, in a weak hyperbolic sense \cite{Nagy,Reula1998}, provided that $T_{ab}$ only depends on the fields and its first derivatives and that the matter equations of motion, Eq.~\eqref{MatterEOM}, are also well-posed. One can show this result using the fact that, the evolution equations for GR with matter, under these assumptions, are quasilinear, diagonal, and second-order hyperbolic, and we can apply Leray's theorem \cite{Leray}, as done in Ref.~\onlinecite{Wald}. Extending this proof to our case, we can conclude that UG has a well-posed initial value formulation as long as neither $T_{ab}$ nor $Q$, which also appears in the dynamical equations, have second derivatives of the fields, and the matter field equations are well posed. Observe that this restriction on the energy-momentum tensor is enough for Eq.~\eqref{4.32 matter} to be a constraint. Of course, these are sufficient conditions; there could be cases that do not meet this hypothesis and still have a well-posed initial value formulation.

\section{Conclusions}\label{conclusions}

One of the most important properties of any theory is its capacity to make predictions out of initial data. Therefore, if UG is to be considered as a viable physical theory, it needs to have a well-posed initial value formulation. Here, we show that UG is well-posed and we find all the constraints. For the latter, we followed the well-known Dirac method; the result is that there are primary constraints, $C_a=0$, and secondary constraints, $S_a=0$. Then, we used these constraints in the evolution equations to cast them in the form of those in GR, which have a well-posed initial value formulation, completing the proposed analysis.

Remarkably, the unimodular constraint does not behave as a constraint in the sense that it is not imposed on the initial data and preserved under evolution. This feature ought to be present in any theory with nondynamical fields like the gravitational sector of the Standard Model Extension \cite{Kostelecky2004}. In the present case, however, the unimodular constraint can always be satisfied by a choice of the lapse function, shedding some light on the role of the UG nondynamical function.

We found it interesting that the equivalence between GR and vacuum UG, in the context of an initial value problem, is not \emph{a priori} obvious (not even when $Q$ is constant). Interestingly, only by requiring dynamical consistency does a spacetime constant emerge (e.g., $R=4\Lambda$ in vacuum) that can be used to show this equivalence. Still, UG can be a suitable test theory to perform numerical computation of modified gravity theories. In addition, our results suggest that one can run GR numerical calculations using only the traceless part of the Einstein equations. Of course, to look for new physical phenomena one needs to consider $\nabla^a T_{ab}\neq 0$.

Finally, given that UG has fewer symmetries than GR \cite{HenneauxTeitelboim}, it has more constraints. However, a detailed counting of constraints is not direct \cite{Henneaux:1989zc}, and, to our knowledge, has only been studied perturbatively \cite{deLeon,Gielen}. Interestingly, UG has been analyzed nonperturbatively with Hamiltonian methods to discuss the problem of time \cite{Unruh,Smolin}, yet, studies of other theories with nondynamical fields \cite{MaratMarco,bonder2023hamiltonian} suggest that a full Hamiltonian analysis is not straightforward.

\begin{acknowledgments}
We acknowledge getting valuable feedback from M. Chernicoff, U. Nucamendi, N. Ortiz, M. Salgado, O. Sarbach, and D. Sudarsky. This project used financial support from CONAHCyT FORDECYT-PRONACES grant 140630 and UNAM DGAPA-PAPIIT grant IN101724.
\end{acknowledgments}

\appendix*

\section{Space and time decomposition}\label{Appendix}

In this Appendix, we discuss fundamental aspects of the $3+1$ formalism we use to describe these theories by evolving geometrical objects, closely following Ref.~\onlinecite[chapter 10.2]{Wald}. Recall that we work assuming that $M$ can be foliated by Cauchy surfaces, $\Sigma_t$, parametrized by a global time function, $t$. We further assume that the normal vector $n^a$ to $\Sigma_t$ is timelike, points to the future,  and is normalized according to $g_{ab}n^a n^b=-1$. On each $\Sigma_t$, the spacetime metric induces a Riemannian metric
\begin{equation}\label{4.1}
h_{ab}=g_{ab}+n_a n_b.
\end{equation}
The inverse of this metric is $h^{ab}=g^{ab}+n^a n^b$ ($g_{ab}$ is the only metric used throughout the text to raise/lower indices). Additionally, $h_{ab} n^a=0$ and $h^a_b$ acts as a projector onto $\Sigma_t$.

Consider an arbitrary spacetime vector field $v^a$. We can express this field as
\begin{equation}\label{4.11}
v^a= v_\perp n^a+ v^a_\parallel,
\end{equation}
where $v^a_\parallel$ is tangential to $\Sigma_t$. When $v^a= v^a_\parallel$, we can think of $v^a$, restricted to $\Sigma_t$, as a vector field on $\Sigma_t$. More generally, a spacetime tensor $\tau^{a_1\cdots a_k}_{b_1\cdots b_l}$ is said to be tangent to $\Sigma_t$ if
\begin{equation}\label{4.13}
\tau^{a_1\cdots a_k}_{b_1\cdots b_l}=h^{a_1}_{c_1}\cdots h^{a_k}_{c_k} h_{b_1}^{d_1}\cdots h_{b_l}^{d_l}\tau^{c_1\cdots c_k}_{d_1\cdots d_l}.
\end{equation}

\begin{figure}[t] 
\centering
\includegraphics[scale=0.35]{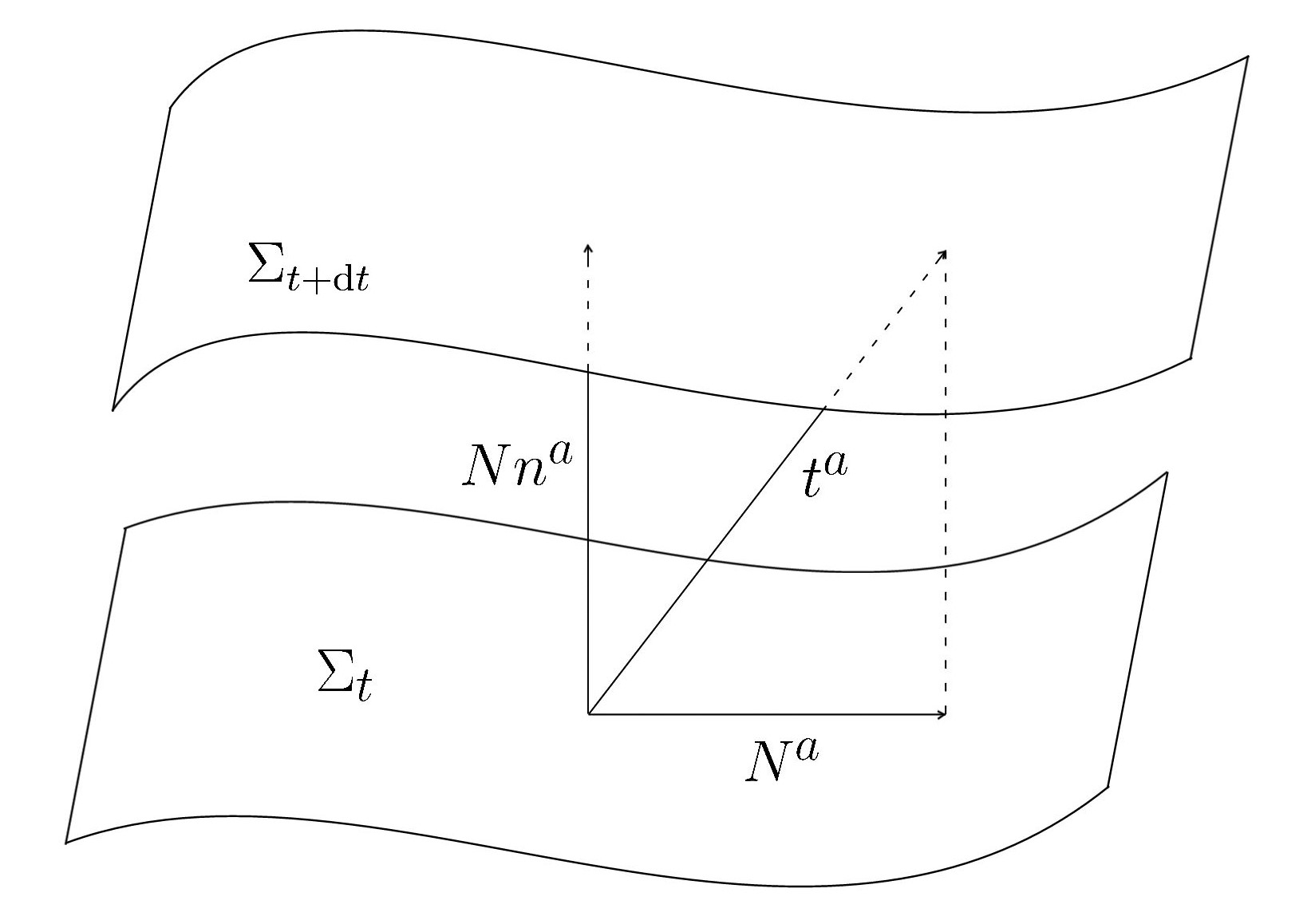}
\caption{A spacetime diagram illustrating the definition of the field $t^a$, the shift vector, $N^a$, and the lapse function, $N$.}
\label{sigma}
\end{figure}

Let $t^a$ be a timelike vector field in $M$ defined by
\begin{equation}\label{4.4}
t^a\nabla_a t=1.
\end{equation}
This vector field identifies points on infinitesimally close hypersurfaces of constant $t$, providing the flow of time used for the evolution. We can decompose $t^a$ into its normal and tangential parts as
\begin{equation}\label{4.5}
t^a= N n^a + N^a,
\end{equation}
where $N$ is the lapse function and $N^a$, which is tangential, is the shift vector. Relevantly, the fact that $t^a$ points to the future implies that $N>0$. Broadly speaking, $N$ gives the rate of change of physical time as compared with $t$. On the other hand, $N^a$ tells us how the coordinates are transported from $\Sigma_t$ to $\Sigma_{t+\mathrm{d}t}$. Fig. \ref{sigma} illustrates this construction.

We define the derivative operator on $\Sigma_t$, $D_a$, by
\begin{equation}\label{4.17}
D_e \tau^{a_1\cdots a_k}_{b_1\cdots b_l}=h^{a_1}_{c_1} \cdots h^{a_k}_{c_k} h_{b_1}^{d_1}\cdots h_{b_l}^{d_l}h_{e}^{f}\nabla_f \tau^{c_1\cdots c_k}_{d_1 \cdots d_l},
\end{equation}
where $\tau^{a_1\cdots a_k}_{b_1\cdots b_l}$ is a tangential tensor. Importantly, one can readily show that this is the only (torsionless) derivative operator such that $D_a h_{bc}=0$ \cite{Wald}. Moreover, the time derivative of any tangential tensor is defined by
\begin{equation}
\dot{\tau}^{a_1\cdots a_k}_{b_1\cdots b_l} = h^{a_1}_{c_1}\cdots h^{a_k}_{c_k}h_{b_1}^{d_1}\cdots h_{b_l}^{d_l} \pounds_t \tau^{c_1\cdots c_k}_{d_1\cdots d_l} ,\label{4.18}
\end{equation}
where $\pounds_t$ is the Lie derivative along $t^a$. Observe that the time derivative is, by construction, a tangential tensor.

We define $K_{ab}\coloneqq h_a^c\nabla_c n_b$, which can be shown to be symmetric. The tensor $K_{ab}$ is known as the extrinsic curvature and it describes the embedding of $\Sigma_t$ in $M$. If we use the time derivative on $h_{ab}$, we obtain
\begin{equation}\label{4.19}
\dot{h}_{ab}= 2N K_{ab}+2D_{(a}N_{ b)},
\end{equation}
showing that $K_{ab}$ is related to the time derivative of $h_{ab}$. We also need some expressions that relate the Riemann tensor associated with $g_{ab}$, $\tensor{R}{_a_b_c^d}$, with objects in $\Sigma_t$. One can show that the purely tangential projection of the Riemann tensor satisfies a Gauss-Codazzi relation \cite{Wald}:
\begin{equation}\label{4.21}
\tensor{R}{_a_b_c^d}h^a_e h^b_f h^c_g h_d^j= \tensor {^{\left(3\right)}R}{_e _f _g ^j}+K_{eg}\tensor{K}{_f ^j}-K_{fg}\tensor{K}{_e ^j} ,
\end{equation}
where $\tensor{^{\left(3\right)}R}{_a _b _c ^d}$ is the Riemann tensor associated with $h_{ab}$. In addition, $\tensor{^{\left(3\right)}R}{_a _b}$ and $^{\left(3\right)}R$, respectively, represent the $3$-dimensional Ricci tensor and curvature scalar. Other projections of the Riemann tensor are
\begin{eqnarray}
\tensor{R}{_a_b_c^d} h^a_e h^b_f h^c_g n_d &=& D_e K_{fg}-D_f K_{eg}\label{4.22} ,\\
\tensor{R}{_a_b_c^d} n^a h^b_e h^c_f n_d&=& h_e^b h_f^c n^a\nabla_a K_{bc}-a_e a_f \nonumber\\
&&-D_{\left(e\right.}a_{\left. f\right)}+\tensor{K}{_e^a}K_{af },\label{4.25}
\end{eqnarray}
where $a^a \coloneqq n^b\nabla_b n^a$ is a tangential vector field. Some useful relations concerning the projections and the trace of the Ricci tensor are given by
\begin{eqnarray}
R_{cd}h^c_a h^d_b &=& {^{\left(3\right)}R_{ab}} +KK_{ab} -2 \tensor {K}{_a^c}K_{cb} +N^{-1}\dot{K}_{ab} \nonumber \\
& & -N^{-1}N^c D_c K_{ab} -2N^{-1}K_{c \left( a \right.}D_{\left.b \right)}N^c \nonumber \\
& & -a_a a_b -D_{\left( a \right.} a_{\left.b \right)},\label{Rhh}\\
R_{cd}n^c n^d &=& -N^{-1} h^{cd} \dot{K}_{cd} + N^{-1} N^c D_c K +a_c a^c \nonumber \\
& & +2 N^{-1 } K^{cd}D_c N_d + K^{cd}K_{cd} +D_c a^c, \nonumber \\
& & \label{Rnn}\\
R_{bc} n^b h^c_a &=& D_b \tensor{K}{^b_a}-D_a K,\label{Rnh}\\
\label{RGC}
R &=& {^{\left(3\right)}}R +K^2 -3K_{cd}K^{cd} +2N^{-1} h^{cd}\dot{K}_{ cd}\nonumber \\
& & -2N^{-1} N^c D_c K -4N^{-1} K^{cd} D_c N_d\nonumber \\
& & -2a_c a^c -2D_c a^c,\label{GC R}
\end{eqnarray}
where $K\coloneqq \tensor{K}{_a^a}$. Also, we can show that
\begin{eqnarray}
\dot{K}_{ef} 
&=& N h_e^b h_f^c n^a\nabla_a K_{bc} +N^a D_a K_{ef}\nonumber\\
&&+ 2K_{a\left(e \right.}D_{\left. f\right)} N^a+ 2N\tensor{K}{_e^a}K_{af}.\label{K dot}
\end{eqnarray}
With this result, Eq.~\eqref{4.25} can be written as
\begin{eqnarray}
\tensor{R}{_a_b_c^d} n^a h^b_e h^c_f n_d&=& N^{-1}\dot{K}_{ef}-N^{-1} N^aD_a K_{ef}\nonumber\\
&& -2N^{-1}K_{a\left(e \right.}D_{\left. f\right)} N^a-\tensor{K}{_e^a}K_{af} \nonumber\\
&& - a_e a_f- D_{\left(e\right.}a_{\left. f\right)}.\label{4.27}
\end{eqnarray}

\bibliography{bibliography.bib} 

\end{document}